\documentclass{elsarticle}
\usepackage{geometry}
\usepackage{nicefrac}

\usepackage{amsmath}
\usepackage{amssymb}

\usepackage{subfigure}

\usepackage{textcomp}
\usepackage[version=3]{mhchem}
\usepackage[utf8]{inputenc}

\usepackage{eurosym}

\usepackage{rotating}
\usepackage{multirow}

\newcommand{\matrixel}[3]{\left< #1 \vphantom{#2#3}\right| #2 \left| #3 \vphantom{#1#2}\right>}
\newcommand{\braket}[2]{\left< #1 \vphantom{#2}\right| \left. #2 \vphantom{#1}\right>}

\newcommand{\di}{\mathrm{d}}
\newcommand{\e}{\,\mathrm{e}}

\newcommand{\pdd}[2]{\frac{\partial #1}{\partial #2}}
\newcommand{\op}[1]{\mathcal{#1}}

\renewcommand{\vec}[1]{{\boldsymbol{{{#1}}}}}

\renewcommand{\imath}{{\mathrm i}}

\makeatletter%
\newcommand{\restoretitle}[1]{
\newcommand{\GNUPLOTspecial}{%
  \@sanitize\catcode`\%=14\relax\special}%
  {\GNUPLOTspecial{"
SDict begin [
  /Title (#1)
  /Subject (#1 Application)
  /Creator (LaTeX)
  /Author (Werner Koch)
  /CreationDate (Mon Mar  5 10:17:40 2012)
  /DOCINFO pdfmark
end
}}%
}
\makeatother

\usepackage[T1]{fontenc}
\usepackage{lmodern}

\usepackage[compact]{titlesec}
\titlespacing*{\chapter}{0cm}{-\topskip}{0.5cm}[0pt]
\titlespacing{\subsubsection}{2pt}{1ex}{0.3ex}

\titleformat{\chapter}{\large \bfseries}{}{1em}{}
\titleformat{\section}{\normalsize \bfseries}{\thesection}{1em}{}
\titleformat{\subsection}{\normalsize \bfseries \itshape}{\thesubsection}{1em}{}
\titleformat{\subsubsection}{\normalsize \bfseries}{\thesubsubsection}{1em}{}

\DeclareUnicodeCharacter{2010}{-}

\begin{document}
\begin{frontmatter}
\title{Wavepacket revivals via complex trajectory propagation}
\author{Werner Koch}
\address{Weizmann Institute of Science, Rehovot, Israel}
\author{David J. Tannor}
\address{Weizmann Institute of Science, Rehovot, Israel}
\begin{keyword}
semiclassics \sep complex trajectories \sep quantum revival
\end{keyword}

\begin{abstract}
Complex-valued semiclassical methods hold out the promise of treating classically allowed and classically forbidden processes on the same footing.  In addition, they provide a natural way to describe optical excitation with complex fields within the trajectory framework. Despite their promise, these methods have until now been limited to short time propagation, due to the numerical difficulties introduced by the complexification.
Using a new Final Value Representation of the Coherent State Propagator (FINCO), combined with an analysis of the complex classical phase space, we achieve accurate wavepacket propagation all the way to the revival time of a strongly anharmonic system.  
\end{abstract}
\end{frontmatter}

\section{Introduction}
Exact quantum mechanical investigation of the dynamics of molecules is to date limited to systems of five or six atoms.
Trajectory based approaches for approximating the time-dependent Schrödinger equation (TDSE) are an attractive alternative for studying large molecular systems.
Among trajectory-based methods, semiclassical initial value representations (SC-IVR), with their conceptual simplicity and high accuracy, have emerged as the most accurate and robust \cite{herman_semiclasical_1984,kay_semiclassical_2004,thoss_semiclassical_2004}.
However, the reliance on real-valued trajectories significantly limits the description of strong quantum effects, e.g. deep tunneling. Moreover, despite impressive work on non-adiabatic transitions and optical excitation using SC-IVR \cite{wu_nonadiabatic_2005,sun_semiclassical_1997}, the ability of real-valued trajectories to handle these processes requires either specialized techniques or the introduction of stochastic elements in the dynamics.

Alternatively, a range of approaches have been developed that employ complex-valued classical trajectories to describe quantum effects, \cite{doll_complexvalued_1973,huber_generalized_1987,boiron_complex_1998,baranger_semiclassical_2001}.
One example of a systematically derived, complex trajectory description of quantum mechanics is Bohmian mechanics with complex action (BOMCA) \cite{goldfarb_bohmian_2006}.
While it has certain similarities to other complex trajectory approaches, it allows for a consistent treatment of dynamical processes including multiple surfaces and optical excitations \cite{zamstein_non-adiabatic_2012,zamstein_non-adiabatic_2012-1,zamstein_complex_2013}.
Furthermore, it was recently shown how a final value representation of the coherent state propagator (FINCO) can be used to eliminate the need for a root search in the complex configuration space \cite{zamstein_communication:_2014}. This paves the way toward more accurate, efficient and robust calculations, akin to the IVR approach in real valued trajectory methods.

Despite their ability in principle to describe strongly quantum phenomena, complex trajectory approaches have been hampered by numerical obstacles introduced by the complexification. As a consequence, results for anharmonic, oscillatory systems until now have been limited to about one period of coherent oscillation \cite{huber_generalized_1988,de_aguiar_initial_2010,zamstein_communication:_2014}.

In this Letter we present semiquantitative wavepacket evolution all the way to the quantum revival time, representing an order of magnitude increase in propagation times for such strongly anharmonic systems. This is achieved by the use of complex time, as well as an analysis of the classical phase space properties of the FINCO propagator.

The underlying theory is summarized in Sec.~\ref{sec:theory} followed by a numerical study of wavepacket revivals in Sec.~\ref{sec:morse}. A summary of the advantages of the FINCO propagator is given in Sec.~\ref{sec:successes} and remaining challenges and unresolved problems are discussed in Sec.~\ref{sec:challenges}.

\section{Theory}
\label{sec:theory}
\subsection{BOMCA }
Following the derivation presented in Refs.~\cite{goldfarb_bohmian_2006,goldfarb_interference_2007}
we insert the ansatz
\begin{align}
\Psi(\tilde{q},t)=\exp\left\{\imath S(\tilde{q},t)\right\}\quad S: \mathbb{C}\times\mathbb{R}\rightarrow\mathbb{C}
\end{align}
into the time-dependent Schrödinger equation (TDSE), obtaining
\begin{align}
\label{eq:TDSE_partial}
\frac{\partial}{\partial t} S =
-V(\tilde{q},t)-\frac{1}{2}(\partial_{\tilde{q}}S)^2+\frac{\imath}{2}\partial_{\tilde{q}}^{2} S\,.
\end{align}
$\Psi(\tilde{q},t)$ is the analytically continued wavefunction depending on complex $\tilde{q}$ and we use atomic units throughout.
Specifying a momentum
\begin{align}
\label{eq:momentum}
  \frac{\di \tilde{q}}{\di t}=\tilde{p}\,,
\end{align} we transform to the moving frame $\frac{\di}{\di t}=\frac{\partial}{\partial t} + \frac{\di \tilde{q}}{\di t}\frac{\partial}{\partial t}$. Taking successive derivatives with respect to
 $\tilde{q}$ yields equations of motion (EOM) for the spatial derivatives $S_n=\frac{\partial^n}{\partial \tilde{q}^n}S(\tilde{q})$
\begin{align}
\label{eq:TDSE_total_n}
\frac{\mathrm{d}}{\mathrm{d}t}S_n&=-V_n-\frac{1}{2}(S_1^2)_{n}+\tilde{p}S_{n+1}+\frac{\imath}{2}S_{n+2}
\end{align}
with $(S_{1}^{2})_{n}=\sum^{n}_{j=0}\binom{n}{j}S_{j+1}S_{n-j+1}$ and
$V_n=\frac{\partial^n}{\partial \tilde{q}^n}V(\tilde{q})$ and the initial conditions
\begin{align}
\label{eq:initial_cond}
S_n(\tilde{q}(t_0),t_0)&=-\imath\left.\pdd{{}^n}{\tilde{q}^n}\ln\left\{\Psi(\tilde{q},t_0)\right\}\right\vert_{\tilde{q}=\tilde{q}(t_0)}\,.
\end{align}
Truncation of the hierarchy of equations~\eqref{eq:TDSE_total_n} at $N$ by setting $S_{N+1}=S_{N+2}=0$, together with Eq.~\eqref{eq:momentum}, results in a closed set of EOMs.

 Specializing to $N=2$ and identifying $\tilde{p}\equiv S_1$, Eqs.~\eqref{eq:momentum} and \eqref{eq:TDSE_total_n} yield classical EOMs for $\tilde{q}(t)$ and $\tilde{p}(t)$ and
\begin{align}
\label{eq:EOM_S0}
\dot{S}_0&=- V_0 +\frac{1}{2}S_1^2+\frac{\imath}{2}S_2\,,\\
\label{eq:EOM_S2}
\dot{S}_2&=-S_2^2-V_2\,.
\end{align}
The terms in Eq.~\eqref{eq:EOM_S0} correspond to the potential, kinetic and quantum parts of the action.

Note that Eqs.~\eqref{eq:EOM_S0} and \eqref{eq:EOM_S2} are equivalent to equations appearing in Generalized Gaussian Wave Packet Dynamics (GGWPD)\cite{huber_generalized_1988,huber_generalized_1987}
\begin{align}
\dot{c}=-\imath V+\frac{\imath}{2}p^2-\imath\alpha,\\
\dot{\alpha}=-2\imath \alpha^2+\frac{\imath}{2}V_2\,,
\end{align}
with $c\equiv \imath S_0$ and $\alpha\equiv -\frac{\imath}{2} S_2$.
However, the initial conditions are not necessarily the same: equations~\eqref{eq:EOM_S0} and \eqref{eq:EOM_S2} can be applied to arbitrary initial states.

Propagation of the entire initial manifold $\tilde{q}(t_0)$ results, in principle, in a representation of the wavefunction $\Psi(\tilde{q}(t),t)$ at all final complex coordinates $\tilde{q}(t;\tilde{q}(t_0))$ at a final time $t$.

\subsection{Reconstruction of the wavefunction}
\label{sec:reconstruction}
In practice, only a finite set of trajectories can be propagated.
This poses the question of how to reconstruct $\Psi(x,t)$ for $x\in\mathbb{R}$. We discuss two straightforward approaches in Sections \ref{sec:root_search} and \ref{sec:analytic_cont}, before turning to the significantly more successful FINCO approach in Section \ref{sec:FINCO}.
\subsubsection{Root search}
\label{sec:root_search}
Starting from a set of initial guesses, a root search can be performed to find all $\tilde{q}(t_0)$ with $\tilde{q}(t;\tilde{q}(t_0))=x\in\mathbb{R}$.
The wavefunction is then recovered from the superposition of all the corresponding contributions
\begin{align}
\label{eq:root_search}
  \Psi(x,t)=\sum\exp\left\{\imath S_0(t)\right\}\,.
\end{align}
The root search may be facilitated by using the stability matrix of the trajectories to yield a Newton-Raphson like approach \cite{goldfarb_bohmian_2006,pal_generalized_2016}.
The resulting wavefunction~\eqref{eq:root_search} can be very accurate but the root search process may be difficult if there are many roots to find.

\subsubsection{Analytic continuation}
\label{sec:analytic_cont}
The root search may be avoided by extrapolating to a given point $x$ on the real axis using the spatial derivatives of the action $S_n(\tilde{q}(t),t)$ from the nearby complex trajectories:

\begin{align}
\label{eq:continuation}
\Psi(x,t)&=\matrixel{x}{\e^{-\imath\op{H}t}}{\Psi(t_0)}\approx \exp\left\{\imath\sum_{n=0}^N\frac{1}{n!}(x-\tilde{q}(t))^nS_n(t)\right\}\,.
\end{align}
This is effectively analytic continuation from $\tilde{q}(t)$ to $x$. Unfortunately, the quality of reconstruction is poor unless the initial manifold sampling is very dense, because the range of validity of the Taylor expansion in the exponent of Eq.~\eqref{eq:continuation} is very limited.

\subsubsection{Gaussian continuation: FINCO}
\label{sec:FINCO}
 A superior method for performing the analytic continuation was introduced in Ref.~\cite{zamstein_communication:_2014} in the form of a final value representation of the coherent state propagator (FINCO). Although we use it here combined with complex trajectory propagation, the method is arguably a new and general method for analytic continuation. As opposed to conventional methods for analytic continuation that use local expansions, this method can traverse large distances in the complex plane by exploiting the non-local equivalence of Gaussian manifolds. The first step in the method is to insert a complete set of coherent states into the analytically  continued Eq.~\eqref{eq:continuation}
\begin{align}
\label{eq:FINCO}
\Psi(x,t)=\frac{1}{2\pi}\int\di p_f\di q_f\braket{x}{g_f}\matrixel{g_f}{\e^{-\imath\op{H}t}}{\Psi(t_0)}.
\end{align}
Here
\begin{align}
\label{eq:gaussian}
  \braket{x}{g_f}=\left(\frac{2\gamma_f}{\pi}\right)^{\frac{1}{4}}\exp\left\{ -\gamma_f\left(x-q_{f}\right)^{2}+\imath p_{f}\left(x-q_{f}\right)\right\}
\end{align}
is a Gaussian state with real valued width, momentum and position parameters $\gamma_f,p_f$ and $q_f$.
Using Eq.~\eqref{eq:continuation} and Eq.~\eqref{eq:gaussian} for $N=2$, the overlap integral $\matrixel{g_f}{\e^{-\imath\op{H}t}}{\Psi(t_0)}$ can be computed explicitly
\begin{align}
\label{eq:overlap}
 \matrixel{g_f}{\e^{-\imath\op{H}t}}{\Psi(t_0)}= \left(\frac{2\gamma_f}{\pi}\right)^{\frac{1}{4}}\sqrt{\frac{\pi}{\gamma_f-\frac{\imath}{2}S_2(t)}}\exp\left\{ \imath S_{0}\left(t\right)
+\gamma_f\left(q_f^2-\tilde{q}(t)^2\right)
+\imath\left(S_1\left(t\right)q_f-p_f\tilde{q}(t)\right)
\right\}\,.
\end{align}

Specializing to Gaussian initial states, $\Psi(t_0)=g_i$, we recognize the overlap matrix element $\matrixel{g_f}{\e^{-\imath\op{H}t}}{\Psi(t_0)}$ as the semiclassical coherent state propagator (SCSP) $\matrixel{g_f}{\e^{-\imath\op{H}t}}{g_i}$.
Normally, the evaluation of the expression for the SCSP, Eq.~\eqref{eq:overlap}, involves a root search from trajectories that evolve from the initial ket manifold at time $t_0$ to the final bra manifold at time $t$.
However, in FINCO the root search is avoided by noting that any complex valued pair of propagated $(\tilde{p}(t),\tilde{q}(t))$ can be identified with a real valued parameter pair $(p_f,q_f)$ through
the Huber-Heller (HH) bra transformation \cite{huber_generalized_1987}
\begin{align}
\label{eq:HH-bra}
  \xi_f=2\gamma_f \tilde{q}(t)-\imath \tilde{p}(t)=2\gamma_f q_f-\imath p_f
\end{align}
in the final manifold. Whatever value of $(\tilde{p}(t),\tilde{q}(t))$ the trajectory ends up with at time $t$ it is a root of the SCSP {\emph{with those final conditions}}. This is similar to shooting the arrow and then drawing the target around it. Exploiting the HH transform, the value of the SCSP with complex center $(\tilde{p}(t),\tilde{q}(t))$ is then used as the coefficient of the Gaussian with real center $(p_f,q_f)$.

Finally, the integration over final phase space variables $(p_f,q_f)$ in Eq.~\eqref{eq:FINCO} is replaced with an integration over initial complex coordinate space variables, $(\Re \tilde{q}(t_0),\Im \tilde{q}(t_0))$. Using the monodromy matrix elements $M_{\alpha\beta}=\frac{\partial\tilde{\alpha}(t)}{\partial\tilde{\beta}(t_0)}$ and the initial manifold condition $\frac{\partial \tilde{p}(t_0)}{\partial \tilde{q}(t_0)}\equiv S_2(t_0)$ for the second derivative of the action, obtained from Eq.~\eqref{eq:initial_cond},
transforms the initial coordinate space variables as $\di p_f\di q_f\rightarrow |J|\di\Re \tilde{q}(t_0)\di\Im \tilde{q}(t_0)$. The determinant of the Jacobian of transformation, $|J|$, is given by
\begin{align}
\label{eq:integration_measure}
|J|=-\frac{1}{2\gamma_f}\left\vert2\gamma_f M_{qq}(t)+2\gamma_f M_{qp}(t)S_2(t_0)-\imath M_{pq}(t)-\imath M_{pp}(t)S_2(t_0)\right\vert^2\,.
\end{align}
A detailed derivation of Eq.~\ref{eq:integration_measure} is given in \ref{app:trafo}.
\ref{app:reformulation} shows how the square root prefactor of Eq.~\eqref{eq:overlap} combined with the exponentiated quantum action can be reformulated more compactly in terms of the Jacobian~\eqref{eq:integration_measure}.

Thus, computing Eqs.~\eqref{eq:gaussian}, ~\eqref{eq:overlap} and ~\eqref{eq:integration_measure} from the entire initial manifold $\tilde{q}(t_0)$ and inserting into Eq.~\eqref{eq:FINCO} yields the desired propagated wavefunction $\Psi(x,t)$
\begin{align}
  \Psi(x,t)=\int\di\Re \tilde{q}(t_0)\di\Im \tilde{q}(t_0)
\frac{-1}{4}\left(\frac{2}{\pi\gamma_f}\right)^{\frac{3}{4}}\left|J\right|^{\frac{3}{4}}
\braket{x}{g_f}\exp\left\{\imath S_{0;\rm{CL}}\left(t\right)
+\gamma_f\left(q_f^2-\tilde{q}(t)^2\right)
+\imath\left(S_1\left(t\right)q_f-p_f\tilde{q}(t)\right)\right\}\,.
\end{align}
with $S_{0;\rm{CL}}$ the classical action.
This equation has certain similarities with  Eq.~(39) in Ref.~\cite{de_aguiar_initial_2010} for the semiclassical coherent state propagator.
The related equation for the wavefunction, Eq.~(37) in Ref.~\cite{de_aguiar_initial_2010}, has a fourfold integral $\di^2z_0 \di^2v_1$ and differs in other particulars from the above equation.

\subsection{Complex time propagation}

Trajectories with complex position and momentum may encounter singularities in the complex plane even if the potential $V(x)$ defined on the real axis is perfectly regular.
In fact, even if the analytically continued potential is singular only at isolated points, and finite time encounters with these points are of measure zero, trajectories that closely approach these points will experience numerical difficulties that will taint them for the future.
Due to the exponential dependence of the wavefunction reproduction on the dynamical quantities of the trajectories, the quality of the final wavefunction reproduction quickly deteriorates if there is any numerical noise.
Moreover, the direction of passage of a trajectory around branch points (dynamical singularities) introduces topological differences that lead to distinctly different solutions and thus to separate contributions to the final wavefunction.

References~\cite{doll_complexvalued_1973,kay_time-dependent_2013,petersen_wave_2015} showed that complex time propagation, along with the correct choice of complex time contours, leads to accurate wavepacket reconstruction for barrier transmission problems. Similarly, the oscillatory system studied in the following section is strongly influenced by the dynamical singularities, and the use of complex time integration contours is required to obtain accurate results.
A brief discussion of the influence of the dynamical singularities can be found in Sec.~\ref{sec:branches} but a detailed analysis is beyond the scope of this Letter.

\section{Numerical example}
\label{sec:morse}
 Consider the Morse oscillator
\begin{align}
V\left(x\right)&=D\left[\left(1-\mathrm{e}^{-\beta x}\right)^{2}-1\right]
\end{align}
with the width parameter $\beta=0.2209$ and the dissociation limit $D=10.25$. These are the same parameters used in previous studies of complex trajectory methods
 \cite{huber_generalized_1987,huber_generalized_1988,zamstein_communication:_2014}.

An initially normalized Gaussian state with width $\gamma_0=\frac{1}{2}$,  central position $q_0=9.342$ and central momentum $p_0=0$ is propagated until the revival time at $t=20 T_{\rm{cl}}$, where $T_{\rm{cl}}\approx 12.88$ is the classical period of a trajectory starting at $q_0,p_0$.
Time slices for the wavefunction reproduced with $\gamma_f=\gamma_0=\frac{1}{2}$ are shown in Fig.~\ref{fig:psi_of_t}.

\begin{figure}[htp]
  \centering
  \begin{minipage}[htp]{0.45\linewidth}
\includegraphics[width=\textwidth]{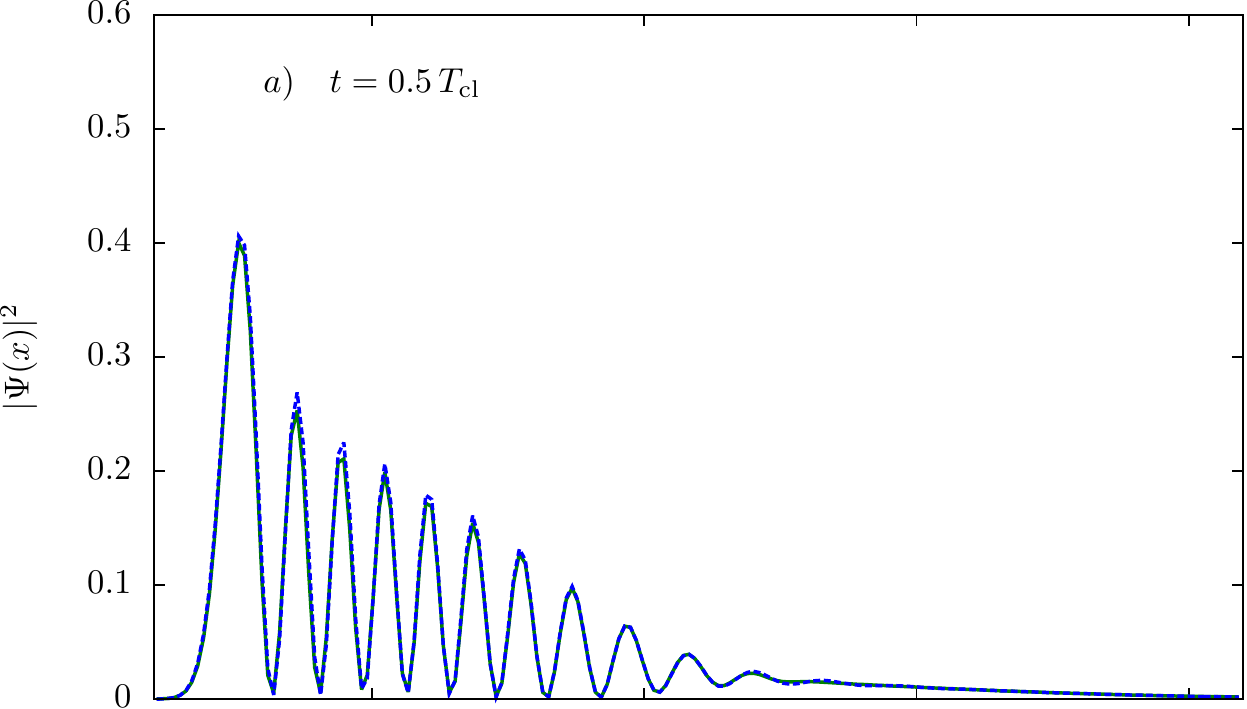}
  \end{minipage}
  \begin{minipage}[htp]{0.45\linewidth}
\includegraphics[width=\textwidth]{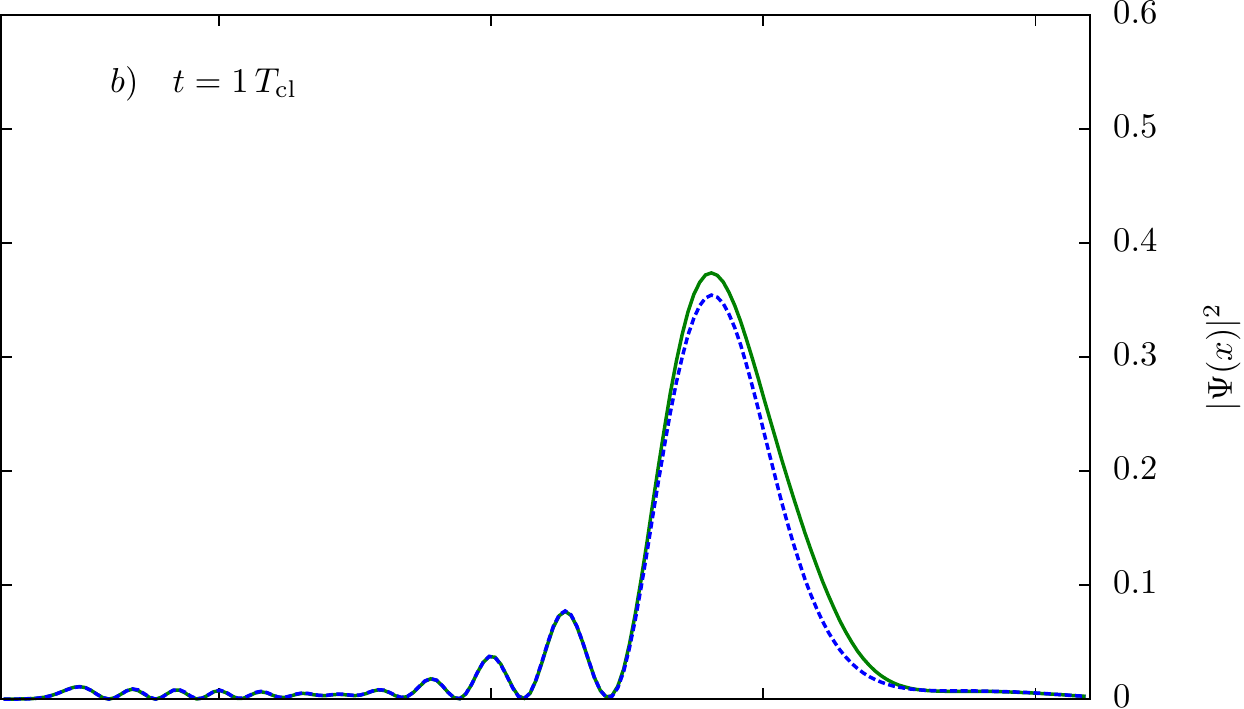}
  \end{minipage}

  \begin{minipage}[htp]{0.45\linewidth}
\includegraphics[width=\textwidth]{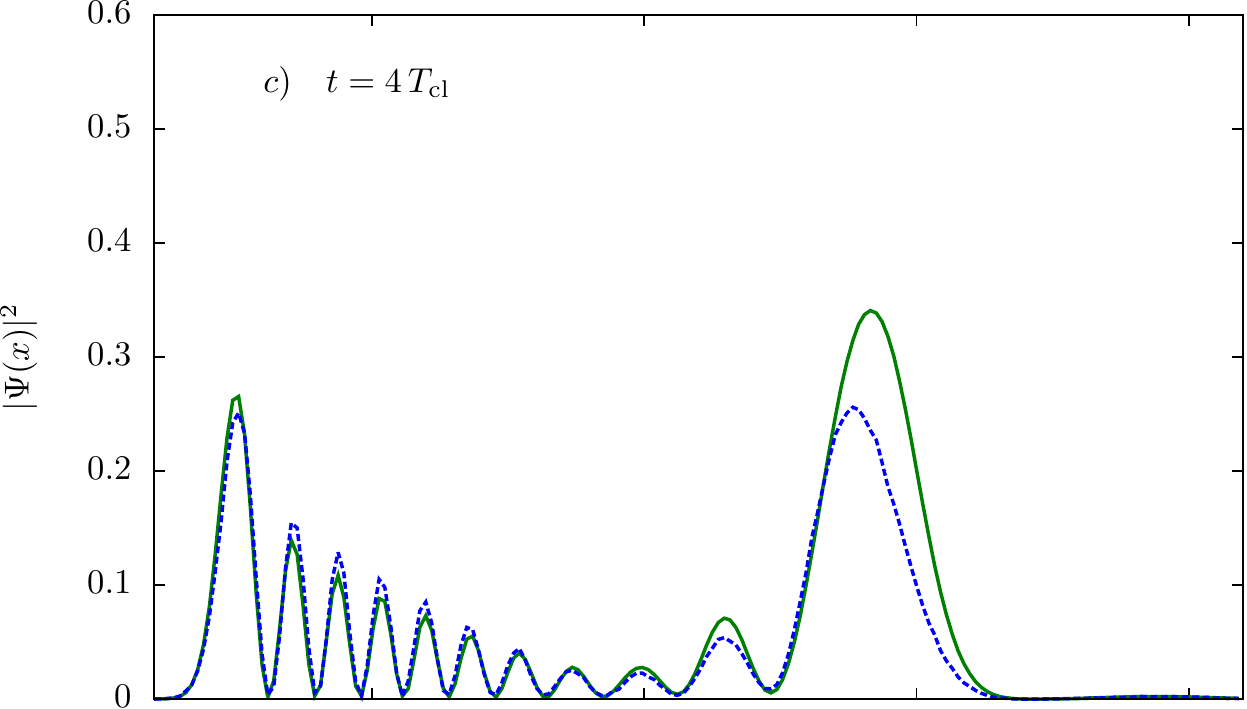}
  \end{minipage}
  \begin{minipage}[htp]{0.45\linewidth}
\includegraphics[width=\textwidth]{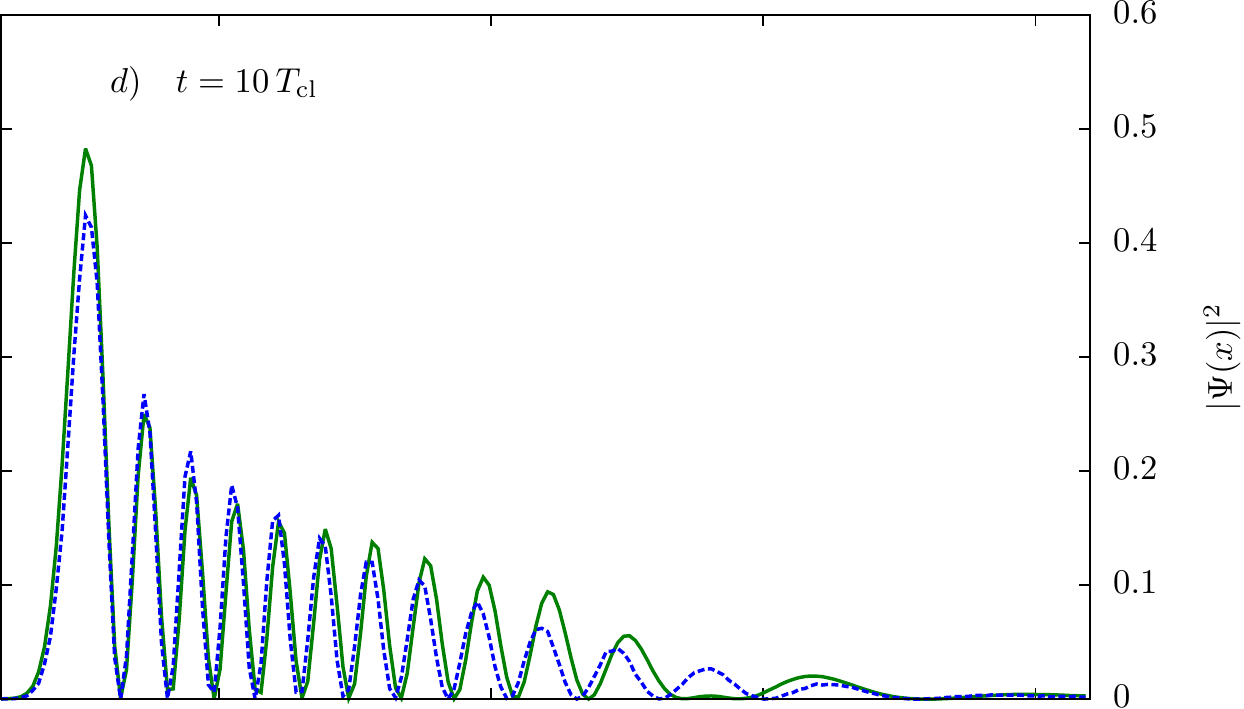}
  \end{minipage}

  \begin{minipage}[htp]{0.45\linewidth}
\includegraphics[width=\textwidth]{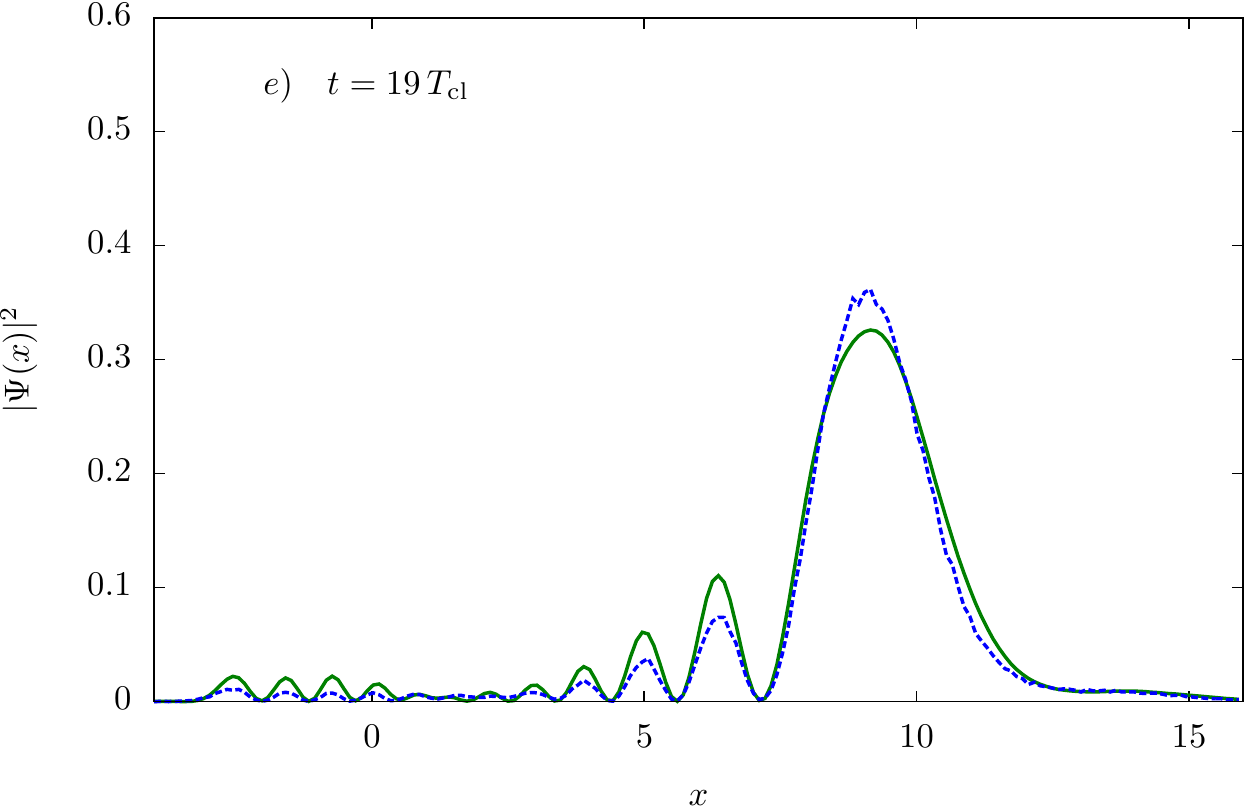}
  \end{minipage}
  \begin{minipage}[htp]{0.45\linewidth}
\includegraphics[width=\textwidth]{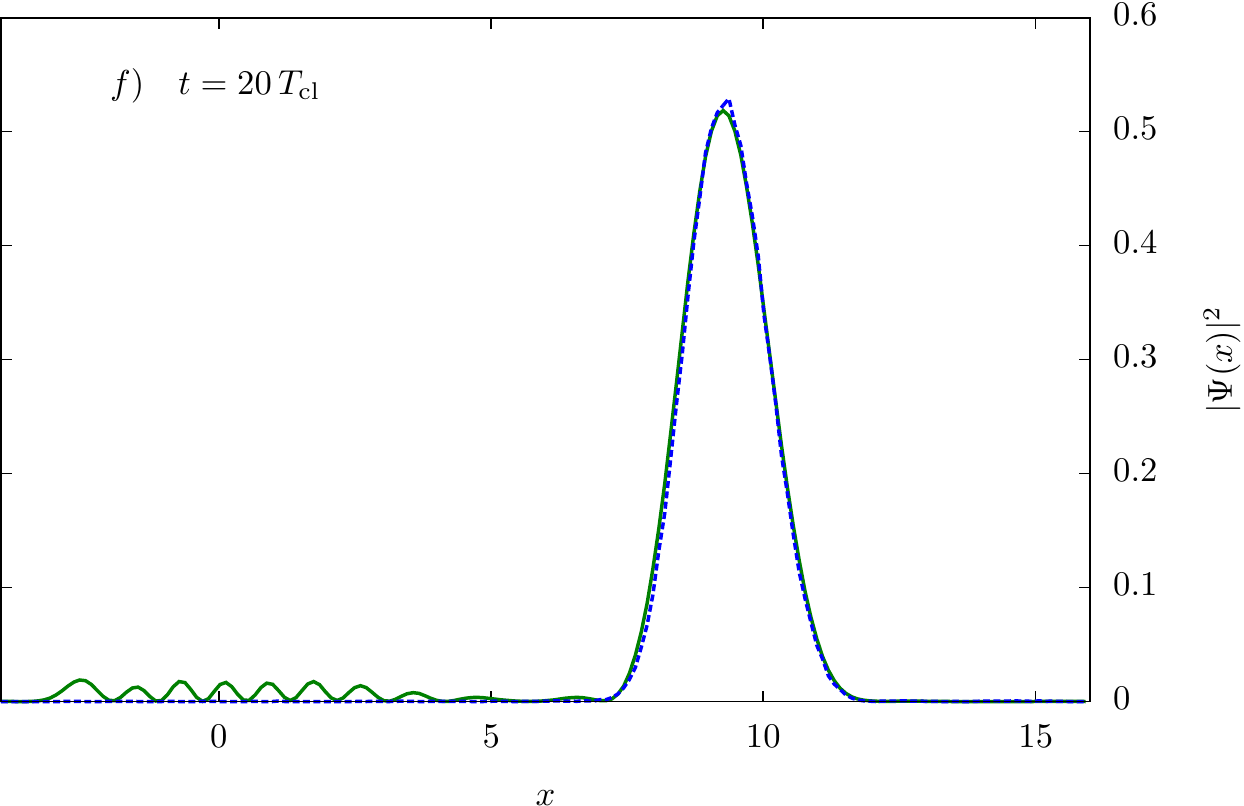}
  \end{minipage}
  \caption{\label{fig:psi_of_t}$|\Psi(x,t)|^2$ from FINCO reconstruction (green solid) and split operator quantum propagation (blue dashed) for a range of times.
Panels a)-f) correspond to 0.5,1,4,10,19,20 classical periods.
Up to 120000 trajectories were used to reconstruct the wavefunctions for the longest time.
}
\end{figure}

The reproduction quality is semi-quantitative until the quantum revival time. The nodal structure of the wavefunction is correctly reproduced all the way until the nineteenth fractional revival. At the revival time of twenty classical periods, the dominant maximum is reproduced well but there are small additional features indicating incomplete destructive interference.

The physical origin of the wavepacket revival is a sum over contributions from trajectories that have completed different numbers of orbits around the minimum of the potential, analytically continued to a saddle point in the complex plane.
This is the analog for complex trajectories of the contributions to the revival from different portions of the filamented classical manifold that encompasses the origin in phase space in real trajectory propagation \cite{suarez_barnes_semiclassical_1993,mallalieu_semiclassical_1994}.
The individual contributions from the complex trajectories loosely correspond to different branches in initial complex coordinate space; see Section \ref{sec:branches}.

The semi-quantitative propagation to twenty classical periods exceeds previously accessible propagation times using complex trajectories by an order of magnitude. The use of complex time is critical to reach these long times, but an understanding of the complex classical phase space is important as well, as will be described in the next section.
Plots of the wavefunction reconstruction without the use of complex time contour can be found in the supplementary material~\ref{sec:real_time}.

\section{Properties of the FINCO propagator in complex classical phase space}
\label{sec:successes}
Because of its similarly to an initial value representation, FINCO benefits from many of the same advantages as real-valued IVRs, but it also has advantages that have no analog in real-valued dynamics. These advantages include a) avoiding a root search, b) removing divergences at caustics, c) automatic summation over multiple roots and d) overcoming internal Stokes phenomena. Advantage a) has already been addressed. We now address advantages b)-d).
\subsection{Behavior at caustics}
\label{sec:caustics}
Caustics are familiar in the context of real-valued trajectory propagation.
In a one-dimensional system, as studied in this Letter, they are places where $\frac{\partial z(t)}{\partial y(t_0)}=0$ where $z(t),y(t_0)$ are any linear combination of final and initial position and momenta respectively\cite{littlejohn_semiclassical_1986}.
(In the multidimensional case, this corresponds to a matrix $\frac{\partial\vec{z}(t)}{\partial\vec{y}(t_0)}$ of less than full rank.)
Geometrically, caustics are a consequence of the projection of the trajectory from phase space to a reduced dimensional space, generally coordinate space. This projection from phase space to coordinate space is present also in complex trajectory calculations, leading to caustics in the submanifold of complex coordinates.

In the present formulation, caustics manifest themselves in the divergence of the action $S_0(t)$ and its second derivative $S_2(t)$.
However, the exponentiated second order expansion of the action in Eq.~\eqref{eq:continuation} remains normalizable through the caustic. Because the function is normalizable, a uniformization scheme such as the overlap integral in FINCO (cf. Eq.~\ref{eq:FINCO}) remains regular at a caustic even if the action $S_0(t)$ of the associated trajectory diverges.

However, as a consequence of the caustic, the second derivative of the action $S_2(t)$ may change sign as the trajectory passes through (or close by) a caustic.
While this would suggest a divergent expansion in Eq.~\eqref{eq:continuation}, the integral Eq.~\eqref{eq:overlap} remains regular as long as $\Re\{\frac{\imath}{2}S_2(t)\}<\gamma_f$.
For the case $\Re\{\frac{\imath}{2}S_2(t)\}>\gamma_f$, the expansion of the complex action up to $N=2$ in momentum space recovers a convergent integral equivalent to Eq.~\eqref{eq:overlap}. However, for the case $\Re\{\frac{\imath}{2}S_2(t)\}\rightarrow\gamma_f$ corresponding to a caustic of $\xi_f=2\gamma_f \tilde{q}(t)-\imath \tilde{p}(t)$, the overlap Eq.~\eqref{eq:overlap} diverges due to the pre-exponential factor.
Fortuitously, this divergence is inconsequential for the FINCO reconstruction since the Jacobian of the transformation from final $(p_f,q_f)$ to initial $(\Re\tilde{q}(t_0),\Im\tilde{q}(t_0))$ space Eq.~\eqref{eq:integration_measure}, exactly cancels this divergence (see \ref{app:reformulation} for details).
This cancellation is analogous to the cancellation of the zero denominator in real-valued IVRs brought about by the Jacobian from final $x$ to initial $p$.

Despite the fact that the divergence of the overlap integral does not prevent the FINCO reconstruction, a phase scar results from it. The implications of this effect are discussed in Sec.~\ref{sec:phase_scar}.

\subsection{Branches in initial coordinate space}
\label{sec:branches}
\begin{figure}[htp]
  \centering
\includegraphics[width=0.7\textwidth]{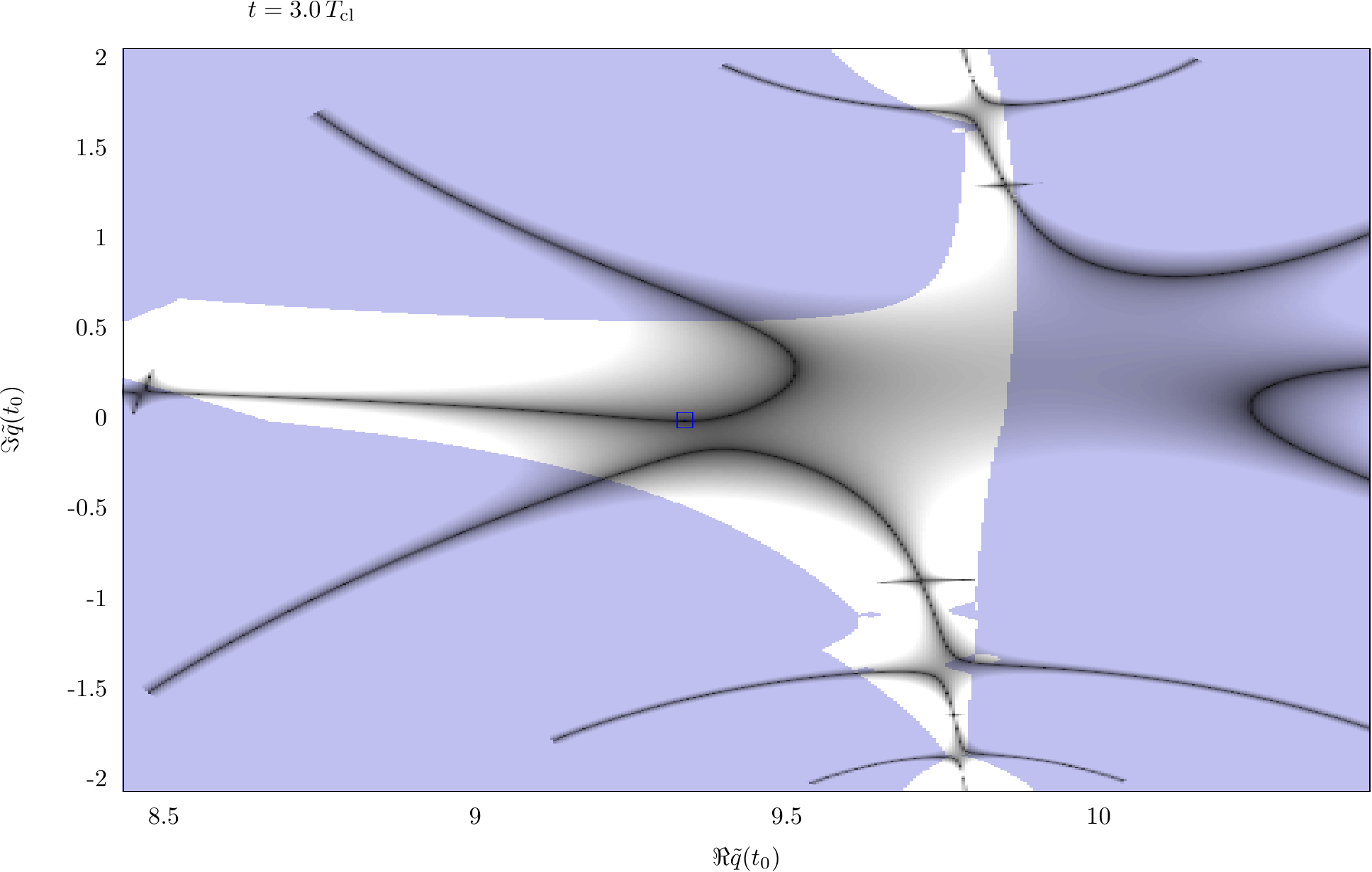}
\caption{\label{fig:branches}Branches of the map $\tilde{q}(t_0)\rightarrow \tilde{q}(t)\in\mathbb{R}$ in coordinate space for $t=3T_{\rm{cl}}$.
Brightness encodes $|\Im \tilde{q}(t;\tilde{q}(t_0))|$, darker denotes closer to the real axis.
Eleven distinct branches are visible.
The blue rectangle at the center highlights the real trajectory corresponding to the center of the Gaussian initial state.
The blue shaded region shows initial conditions for which a purely real time contour is insufficient.
}
\end{figure}

The propagated wavefunction shown in Fig.~\ref{fig:psi_of_t}a)-f) in principle contains contributions from all of initial coordinate space.
A root search along the lines of Sec.~\ref{sec:root_search} would have to account for contributions from all branches of initial coordinates that reach the real axis at time $t$.
The number of branches to account for depends on the propagation time $t$ and for the Morse system increases by four or five per classical oscillation period.
All branches terminate at an internal singularity on one end and run to infinity on the other.
The eleven significant branches for $t=3T_{\rm{cl}}$ are shown in Fig.~\ref{fig:branches}.
Figure \ref{fig:branches} also shows that only part of the relevant branches in initial coordinate space are accessible with a purely real time contour.
At the revival time $t=20T_{\rm{cl}}$ there are 104 significantly contributing branches (not shown).
For longer propagation times, a growing fraction of the trajectories requires complex time contours to circumnavigate the dynamical singularities on the correct side.
This corresponds to the effect in real time propagation of an ever growing fraction of the initial manifold that drifts farther and farther into the complex plane, never to return to the real $x$-axis and contribute to the final wavepacket.

\begin{figure}[htp]
  \centering
\includegraphics[width=\textwidth]{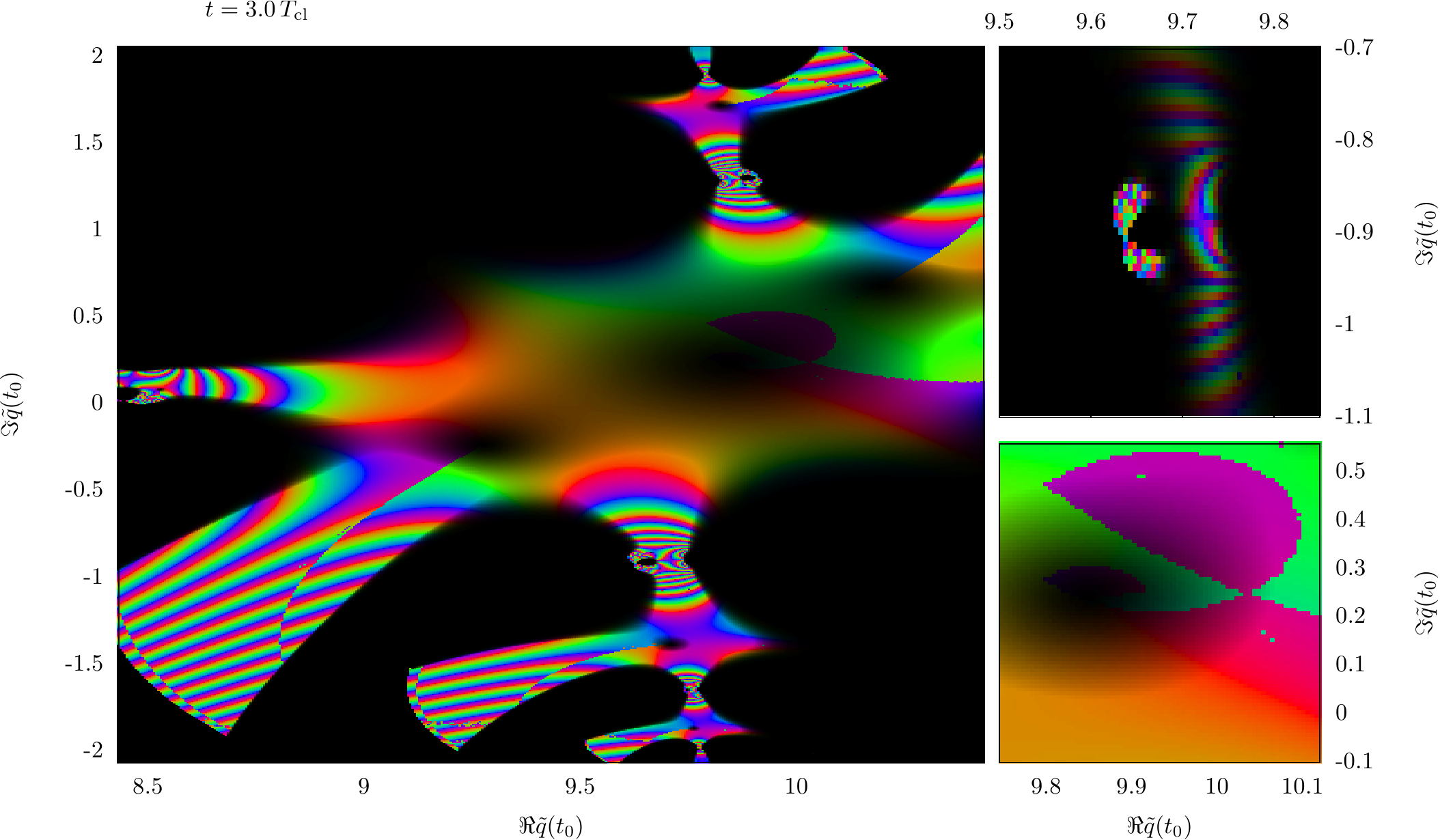}
\caption{\label{fig:ovl} The contributions to the FINCO integral $|J|\left<g_f|\e^{-\imath\op{H}t}|\Psi(t_0)\right>$
 in initial coordinate space for $t=3T_{\rm{cl}}$.
Brightness encodes magnitude (brighter is larger, black is non-contributing) and color (online) denotes phase.
The largest contributions are along the branches of Fig.~\ref{fig:branches}.
The outsets show magnified sections of the initial coordinate space with the magnitude (brightness) rescaled to more clearly show the following features:
A non-contributing caustic of $\xi_f=2\gamma_f \tilde{q}(t)-\imath \tilde{p}(t)$ (see Sec.~\ref{sec:caustics}) with a phase scar emanating from it (see Sec.~\ref{sec:phase_scar}) can be observed at $\tilde{q}(t_0)=9.87+0.25\imath$ (lower outset).
A regularized divergence at a singularity (see Sec.~\ref{sec:internal_stokes}) is located at $\tilde{q}(t_0)=9.64-0.90\imath$ (upper outset).
Divergences related to the kinetic action in the lower left and upper right quadrant (main figure) and related to divergent regions of the potential near $\tilde{q}(t_0)=9.63-0.86\imath$ and $\tilde{q}(t_0)=9.64-0.93\imath$ (upper outset) are discussed in Sec.~\ref{sec:divergences}.
}
\end{figure}

FINCO automatically integrates across all of these branches.
The overlap integrals associated with each trajectory weighted by the integral measure is shown in Fig.~\ref{fig:ovl}.
For long propagation times, it is efficient to sample adaptively from all regions of the initial coordinate space having different numbers of orbits around the minimum of the potential.

\subsection{Internal Stokes' phenomenon}
\label{sec:internal_stokes}
Except for the real branch, i.e. the branch emanating from the center of the initial Gaussian wavepacket
that contains the real trajectory with $\tilde{q}(t_0)=q_0,\tilde{p}(t_0)=p_0$, the classical action diverges in some region of each of the branches discussed in Sec.~\ref{sec:branches}.
Half of these branches diverge close to the singularities
in the coordinate space distribution.

These divergences have been discussed in Ref.~\cite{huber_generalized_1988}.
Reconstruction of the wavefunction by means of a root search as in Sec.~\ref{sec:root_search} can deal with this problem through the tedious procedure of analyzing the underlying structure of the Stokes' phenomenon \cite{shudo_stokes_1996}.
FINCO regularizes these divergences near singular points of the overlaps in Eq.~\eqref{eq:overlap} through the counter term
$\imath\left(S_1\left(t\right)q_f-p_f\tilde{q}(t)\right)$
as the divergences invariably coincide with large imaginary final momenta $\tilde{p}(t)\equiv S_1(t)$.

\section{Remaining challenges}
Despite the significant progress in understanding the properties of FINCO propagation, there are a few aspects of the complex trajectory dynamics that we still do not fully understand.

\label{sec:challenges}
\subsection{Divergence of classical dynamics far in the complex plane}
\label{sec:divergences}
In addition to the internal Stokes' phenomenon near the singular points of the initial coordinate distribution there are two more types of divergences.

One is complementary to the divergence discussed in Sec.~\ref{sec:internal_stokes} but is located towards the exterior of the initial coordinate space distribution.
Empirically this correlates with trajectories for which the imaginary part of the kinetic action $S_{\rm{kin}}(t)=\int_{t_0}^{t}\di t' \frac{1}{2} \tilde{p}(t')^2$ assumes a large negative value.
Besides it being a manifestation of Stokes' phenomenon caused by the approximation in the present approach, it has a dynamical manifestation in the form of a slight asymmetry of the imaginary part of the classical action with respect to the turning points.
For the classical dynamics in complex space this is inconsequential, but for the reconstruction of the wavefunction, this asymmetry leads to vanishing contributions at the extremes of two quadrants of initial coordinate space and exploding contributions in the other two.
Until a more rigorous treatment in terms of the Stokes' phenomenon can be formulated, we simply remove trajectories for which $\Im S_{\rm{kin}}(t)<\sigma$ for a given time $t$.
For the Morse system in Sec.~\ref{sec:morse} the value chosen is $\sigma=-2.5$.

A similar, but different divergence is found for trajectories that stray too far away from the real axis in final coordinate space.
As the real part of the potential becomes negative and thus unbound for large $\Im\tilde{q}$, trajectories venturing into these regions become unphysical.
Since they appear to emerge unscathed at later times we again simply remove trajectories during the interval for which $\Re V(\tilde{q}(t))<\nu$.
For the Morse system in Sec.~\ref{sec:morse} the value chosen is $\nu=-20$.
We note in passing that no such divergences have been found in the Coulomb or Eckart systems.

Due to the exponential dependence of the reconstruction on the dynamical quantities of the trajectories, the results presented in Sec.~\ref{sec:morse} are sensitive to the values of $\nu$ and $\sigma$ to a certain extent.
However, varying either of them by a factor of two up or down will locally (in $x$) affect the magnitude of $\Psi(x,t)$ but preserve the nodal structure.
The value of $\nu$ mostly influences $\Psi(x,t)$ on the steep side of the potential whereas the value of $\sigma$ mostly influences $\Psi(x,t)$ on the shallow side of the potential.

\subsection{Residual numerical noise}
\label{sec:noise}
With the current implementation there is some residual numerical noise in the trajectories that leads to sporadic but substantial deviations.
Removing these trajectories can easily be done with an absolute limit for $\left||J| \matrixel{g_f}{\e^{-\imath\op{H}t}}{\Psi(t_0)}\right|<\epsilon$ with $\epsilon=10^4$.
This procedure does not strongly depend on the value of $\epsilon$ over a few orders of magnitude and only two of the reproductions in Sec.~\ref{sec:morse} required such a removal, affecting one and two trajectories respectively.

\subsection{Emergence of phase scars}
\label{sec:phase_scar}
The most puzzling and so far unresolved problem is an ambiguity of phase of the overlap in Eq.~\eqref{eq:overlap}.
In the presented form this ambiguity results from the multivalued nature of the square root prefactor.
Traditionally, such prefactors are required to be continuous in time \cite{huber_generalized_1988}.
For the similar Herman-Kluk prefactor, a log-derivative reformulation in Ref.~\cite{gelabert_log-derivative_2000} was claimed to be free of phase jumps entirely, obviating checks for continuity.
The latter, however can only be the case for real trajectories.
For complex trajectories, any loop in coordinate space around a caustic must contain a phase jump, irrespective of the definition of the phase.

Worse still, ``continuity in time'' is not well-defined if complex time contours are involved. Continuity can be maintained only along a given contour, but any two contours in complex time with identical initial and final times but enclosing a caustic will incur a phase difference at the final time.

The effect of this ambiguity can be seen in Fig.~\ref{fig:ovl} as an isolated purple region around $\tilde{q}(t_0)=10+0.5\imath$, discontinuously embedded in a green region, and a line of discontinuity emanating from it towards positive real infinity.
Upon close inspection, a similar jump can be found emanating from each of the internal dark spots, corresponding to caustics of the HH bra transformation~\eqref{eq:HH-bra} and thus divergences of the prefactor in Eq.~\eqref{eq:overlap}.
The isolated region can be eliminated by a modified choice of the complex time integration contour, but the discontinuity towards infinity can not.
The latter is intrinsic to the present approach and any semiclassical scheme like it.

For the Morse oscillator system, these phase ambiguities are not overly problematic, although they likely contribute to the imperfect rephasing at the revival time observed in Fig.~\ref{fig:psi_of_t}.
For other systems, however, they preclude reliable reconstruction as they appear right in the center of the most important regions of coordinate space. Even in these problematic cases the trajectory data is sufficient to accurately reconstruct the wavefunction, but further improvements to the present method are required to make the  reconstruction prescriptive.

\section{Conclusions}

In summary, we have demonstrated long time, semi-quantitative wavepacket evolution in a strongly anharmonic system using FINCO, a complex trajectory based propagation scheme. The propagation out to wavepacket revival time, 20 periods of the Morse oscillator, exceeds previously accessible propagation times using complex trajectory methods by an order of magnitude.

The Final Value Representation of the Coherent State Propagator method, FINCO, solves several difficulties with complex-valued trajectory methods: a) there is no need for a root search, b) divergences at caustics are removed, c) contributions from multiple roots are automatically summed and d) internal Stokes phenomena present no difficulty.
While a few unresolved problems remain in the present approach, the quality of reproduction and stability of the approach are now sufficient to pursue a wide variety of new applications. We now describe a few of the key ones.

We have previously shown that BOMCA, the complex trajectory precursor to FINCO, can handle nonadiabatic dynamics very naturally and very accurately Ref.~\cite{zamstein_non-adiabatic_2012,zamstein_non-adiabatic_2012-1}.  The derivation of the EOM from the TDSE is simple and rigorous. It leads to a formulation where there is no surface hopping; trajectories on different potential energy surfaces communicate with each other and exchange amplitude through the difference in their complex phase.

It is a short step from nonadiabatic processes to optical excitation with ultrashort laser pulses, by replacing the off-diagonal coupling $V_{ab}$ with the laser field times transition dipole moment, $-\mu_{ab}E(t)$ \cite{zamstein_complex_2013}.
This should be of particular interest to the ultrafast community in that it provides the ability to incorporate the time dependent laser excitation into trajectory calculations.
Although there has been some very important work along these lines \cite{mitric_laser-field-induced_2009,richter_sharc:_2011}, we believe that the complex-trajectory formulation presented here is the ideal method to treat phase- and amplitude-shaped laser pulses and trajectory propagation within a single consistent framework.
Preliminary calculations incorporating laser pulse sequences into BOMCA were only partially successful; we are interested in returning to these calculations with the greatly enhanced capabilities of FINCO.

Another fascinating possibility is to apply FINCO to study attosecond electron dynamics and high harmonic generation. High harmonics generation has previously been studied with the Herman-Kluk propagator in Ref.~\cite{van_de_sand_irregular_1999,zagoya_dominant-interaction_2012}, but because the HK propagator uses only real-valued trajectories this study was limited to the classically allowed evolution of the electron in the presence of the Coulomb and laser fields.
The potential with FINCO to treat the classically forbidden process of strong field tunneling ionization on a consistent footing with the classically allowed propagation would provide a powerful conceptual and numerical addition to the theoretical tools for studying attosecond physics.
In recent years there has been great interest in the complex time and complex momentum that correspond to the initial conditions for electron trajectories whose recollision leads to high harmonic generation \cite{kaushal_nonadiabatic_2013,pedatzur_attosecond_2015}; it will be very interesting to see if these correspond to the complex trajectories that will emerge in FINCO.

Finally, now that the need for a root search has been eliminated and long time propagation is accurate and stable, we hope that complex trajectory methods will prove useful for studying strong quantum effects in multidimensional systems that cannot be accessed by other theoretical techniques.

\vspace{1EM}
We would like to thank Kenneth G. Kay for valuable discussions concerning the complex classical dynamics and complex time contour propagation. This work was supported by the Israel Science Foundation (533/12 and 1094/16), the Minerva Foundation with funding from the Federal German Ministry for Education and Research and the German-Israeli Foundation for Scientific Research and Development (GIF).

\appendix

\section{Integration measure for FINCO}
\label{app:integration_measure}

\subsection{Transformation from $(p_f,q_f)$ to $(\Re\tilde{q}(t_0),\Im\tilde{q}(t_0))$}
\label{app:trafo}

By virtue of the HH-bra transformation
\begin{align}
\label{eq:HH-bra-repeat}
  \xi_f=2\gamma_f \tilde{q}(t)-\imath \tilde{p}(t)=2\gamma_f q_f-\imath p_f\,,
\end{align}
the Gaussian identity integral is changed from the real parameters $(p_f,q_f)$ to the real and imaginary component of $\xi_f$
\begin{align}
  \di p_f\di q_f=-\frac{1}{2\gamma_f}\di\Re\xi_f\di\Im\xi_f\,.
\end{align}
Changing to initial coordinate space results in
\begin{align}
-\frac{1}{2\gamma_f}\di\Re\xi_f\di\Im\xi_f=-\frac{1}{2\gamma_{f}}\left|J\right|\di{\Re\tilde{q}\left(t_{0}\right)}\di{\Im\tilde{q}\left(t_{0}\right)}\,,
\end{align}
with the determinant of the Jacobian of transformation
\begin{align}
\label{eq:j1}
\left|J\right|&=\pdd{\Re\xi_{f}}{\Re\tilde{q}\left(t_{0}\right)}\pdd{\Re\xi_{f}}{\Re\tilde{q}\left(t_{0}\right)}+\pdd{\Im\xi_{f}}{\Re\tilde{q}\left(t_{0}\right)}\pdd{\Im\xi_{f}}{\Re\tilde{q}\left(t_{0}\right)}\\
\label{eq:j2}
&=\left|\pdd{\Re\xi_{f}}{\Re\tilde{q}\left(t_{0}\right)}+\imath\pdd{\Im\xi_{f}}{\Re\tilde{q}\left(t_{0}\right)}\right|^{2}=\left|\pdd{\xi_{f}}{\Re\tilde{q}\left(t_{0}\right)}\right|^{2}=\left|\pdd{\xi_{f}}{\tilde{q}\left(t_{0}\right)}\right|^{2}\\
\label{eq:Jac-preliminary}
&=\left|\pdd{\xi_{f}}{\tilde{q}\left(t\right)}\left(\pdd{\tilde{q}\left(t\right)}{\tilde{q}\left(t_{0}\right)}+\pdd{\tilde{q}\left(t\right)}{\tilde{p}\left(t_{0}\right)}\pdd{\tilde{p}\left(t_{0}\right)}{\tilde{q}\left(t_{0}\right)}\right)+\pdd{\xi_{f}}{\tilde{p}\left(t\right)}\left(\pdd{\tilde{p}\left(t\right)}{\tilde{q}\left(t_{0}\right)}+\pdd{\tilde{p}\left(t\right)}{\tilde{p}\left(t_{0}\right)}\pdd{\tilde{p}\left(t_{0}\right)}{\tilde{q}\left(t_{0}\right)}\right)\right|^{2}\,.
\end{align}
In passing from Eq.~\eqref{eq:j1} to Eq.~\eqref{eq:j2} we have used the Cauchy-Riemann equations.

Using the second half of Eq.~\eqref{eq:HH-bra-repeat}, the monodromy matrix elements $M_{\alpha\beta}=\frac{\partial\tilde{\alpha}(t)}{\partial\tilde{\beta}(t_0)}$ and the initial manifold condition $\frac{\partial \tilde{p}(t_0)}{\partial \tilde{q}(t_0)}\equiv S_2(t_0)$ for the second derivative of the action, obtained from Eq.~\eqref{eq:initial_cond},
the transformation to initial coordinate space is thus written as
\begin{align}
\label{eq:jacobian}
\di p_f \di q_f=-\frac{1}{2\gamma_f}\left|2\gamma_f M_{qq}(t)+2\gamma_f M_{qp}(t)S_2(t_0;\tilde{q}(t_0))-\imath M_{pq}(t)-\imath M_{pp}(t)S_2(t_0;\tilde{q}(t_0))\right|^2\di\Re \tilde{q}(t_0)\di\Im \tilde{q}(t_0)\,.
\end{align}
For a Gaussian initial state, $S_2(t_0;\tilde{q}(t_0))=2\imath\gamma_0$ is constant.
If furthermore $\gamma_f=\gamma_0$, the Jacobian is equivalent to an expression derived in Refs.~\cite{de_aguiar_initial_2010,kay_jacobian_2013}.
For $\gamma_f\neq\gamma_0$ the final form is different and only Eq.~\eqref{eq:jacobian} leads to a correct reconstruction.

\subsection{Jacobian reformulation in terms of the preexponential factor of the overlap integral}
\label{app:reformulation}

The FINCO overlap integral in Eq.~\eqref{eq:overlap} diverges whenever $\Re\{\frac{\imath}{2}S_2(t)\}\rightarrow\gamma_f$.
It can be shown, however, that this does not lead to a divergence of the wavefunction reconstruction because the Jacobian of transformation derived in \ref{app:trafo} regularizes this divergence.

Following Ref.~\cite{heller_classical_1976}, the equation of motion for $S_2$ may be reformulated as two coupled EOMs for $P_z$ and $Z$ with $S_2=\frac{P_z}{Z}$
\begin{subequations}
\label{eq:EOM_Pz_Z}
\begin{align}
\dot{P}_{z}&=-V_2\left(\tilde{q}\right)Z\\
\dot{Z}&=P_z\\
\label{eq:EOM_Pz_initial}
P_z(t_0)&=S_2(t_0)\\
\label{eq:EOM_Z_initial}
Z(t_0)&=1\,.
\end{align}
\end{subequations}
This substitution allows for the quantum action to be integrated analytically
\begin{align}
  \dot{S}_{0;\rm{QM}}&=\frac{\imath}{2}S_2=\frac{\imath}{2}\frac{P_z}{Z}\\
S_{0;\rm{QM}}(t)&=\frac{\imath}{2}\int_{t_0}^{t}\di t'\frac{P_z}{Z}=\frac{\imath}{2}\int_{t_0}^{t}\di t'\frac{\dot{Z}}{Z}=\frac{\imath}{2}\ln\left(\frac{Z(t)}{Z(t_0)}\right)\,.
\label{eq:Sqm}
\end{align}
Note that the ambiguity of phase discussed in Sec.~\ref{sec:phase_scar} appears in Eq.~\eqref{eq:Sqm} in the choice of integration contour for the integral.

Inserting Eq.~\eqref{eq:Sqm} into Eq.~\eqref{eq:overlap} results in
\begin{align}
 \matrixel{g_f}{\e^{-\imath\op{H}t}}{\Psi(t_0)}&= \left(\frac{2\gamma_f}{\pi}\right)^{\frac{1}{4}}\sqrt{\frac{\pi}{\gamma_f-\frac{\imath}{2}\frac{P_z(t)}{Z(t)}}}\exp\left\{-\frac{1}{2}\ln\left(\frac{Z(t)}{Z(t_0)}\right)+\sigma\right\}\\
&=\left(\frac{2\gamma_f}{\pi}\right)^{\frac{1}{4}}\sqrt{\frac{\pi Z(t_0)}{\gamma_fZ(t)-\frac{\imath}{2}P_z(t)}}\exp\left\{\sigma\right\}\,,
\label{eq:overlap_Pz_Z}
\end{align}
with $\sigma= \imath S_{0;\rm{CL}}\left(t\right)
+\gamma_f\left(q_f^2-\tilde{q}(t)^2\right)
+\imath\left(S_1\left(t\right)q_f-p_f\tilde{q}(t)\right)$ and $S_{0;\rm{CL}}$ the remaining classical action.
Noting that $P_z$ and $Z$ follow the same linear EOMs as the monodromy matrix elements
\begin{align}
  \frac{\di}{\di t}\left(
    \begin{matrix}
      M_{pp}&M_{pq}\\
      M_{qp}&M_{qq}
    \end{matrix}\right)=\left(
    \begin{matrix}
      0&&-V_2\\
      1&&0
    \end{matrix}\right)\left(
    \begin{matrix}
      M_{pp}&M_{pq}\\
      M_{qp}&M_{qq}
    \end{matrix}\right)\,,
\end{align}
and using the initial conditions Eqs.~\eqref{eq:EOM_Pz_initial} and~\eqref{eq:EOM_Z_initial}, their values at time $t$ can be obtained as
\begin{align}
\label{eq:Pz_Z_t}
  \left(
    \begin{matrix}
P_z(t)\\
Z(t)
    \end{matrix}
\right)
=\left(
    \begin{matrix}
      M_{pp}(t)&M_{pq}(t)\\
      M_{qp}(t)&M_{qq}(t)
    \end{matrix}\right)\left(    \begin{matrix}
S_2(t_0)\\
1
    \end{matrix}
\right)\,.
\end{align}
Inserting Eqs.~\eqref{eq:EOM_Z_initial} and~\eqref{eq:Pz_Z_t} into Eq.~\eqref{eq:overlap_Pz_Z} results in
\begin{align}
\label{eq:overlap_monodromy}
\matrixel{g_f}{\e^{-\imath\op{H}t}}{\Psi(t_0)}&=\left(\frac{2\gamma_f}{\pi}\right)^{\frac{1}{4}}\sqrt{\frac{2\pi }{2\gamma_f M_{qq}(t)+2\gamma_f M_{qp}(t)S_2(t_0)-\imath M_{qp}(t)-\imath M_{pp}(t)S_2(t_0)}}\exp\left\{\sigma\right\}\,.
\end{align}

The denominator of the square root prefactor in Eq.~\eqref{eq:overlap_monodromy} is identical to the Jacobian prefactor in Eq.~\eqref{eq:jacobian} and inserting Eqs.~\eqref{eq:jacobian} and \eqref{eq:overlap_monodromy} into Eq.~\eqref{eq:FINCO} yields for the FINCO wavefunction reconstruction
\begin{align}
  \Psi(x,t)=\int\di\Re \tilde{q}(t_0)\di\Im \tilde{q}(t_0)
\frac{-1}{4}\left(\frac{2}{\pi\gamma_f}\right)^{\frac{3}{4}}\left|J\right|^{\frac{3}{4}}
\braket{x}{g_f}\exp\left\{\sigma\right\}\,.
\end{align}
Any divergence in the square root prefactor of the FINCO overlap integral is thus regularized by the vanishing Jacobian.

\section{Real time reconstruction}
\label{sec:real_time}
To illustrate the need for complex time contour propagation, a comparison with a wave function reconstruction from trajectories that do not require a strictly complex time contour for them to contribute significantly is presented in Fig.~\ref{fig:psi_of_t}.

With purely real time propagation, starting at about half a classical period, significant regions of initial coordinate space will result in trajectories that pass beyond a branching point and consequently drift far away from the real axis, never to return.

\begin{figure}[htp]
  \centering
  \begin{minipage}[htp]{0.45\linewidth}
\includegraphics[width=\textwidth]{"finco2D_wf_0_5"}
  \end{minipage}
  \begin{minipage}[htp]{0.45\linewidth}
\includegraphics[width=\textwidth]{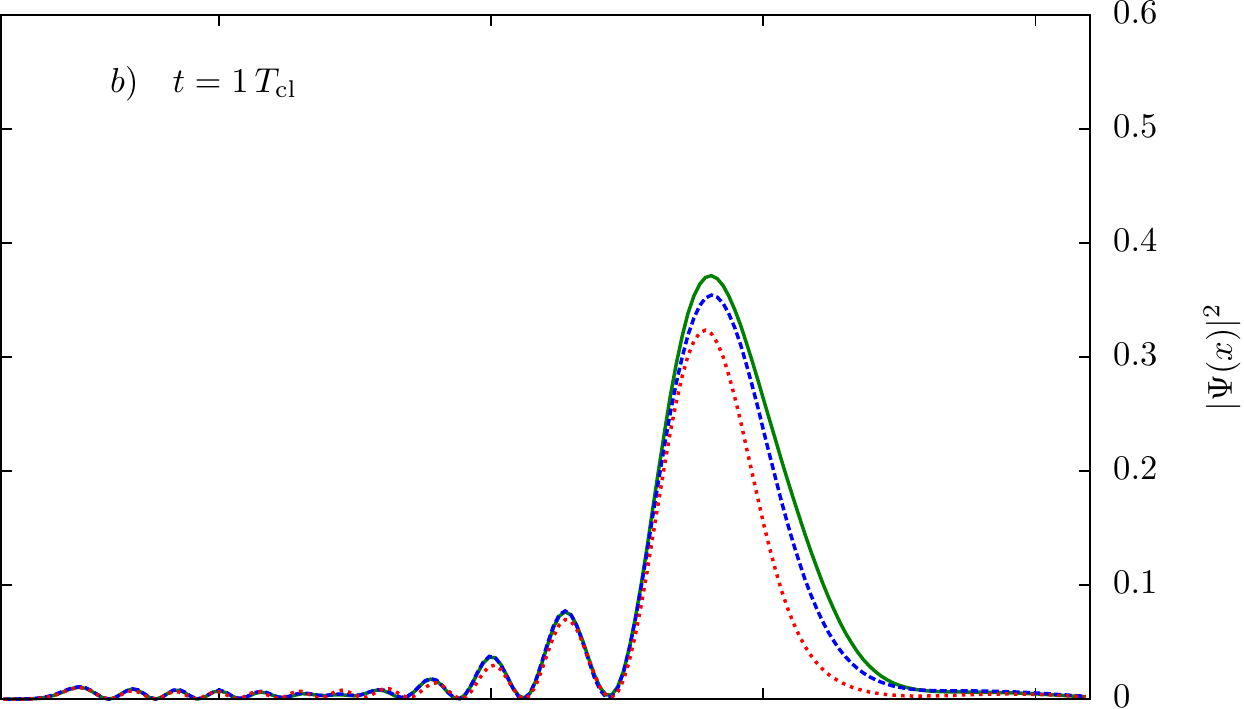}
  \end{minipage}

  \begin{minipage}[htp]{0.45\linewidth}
\includegraphics[width=\textwidth]{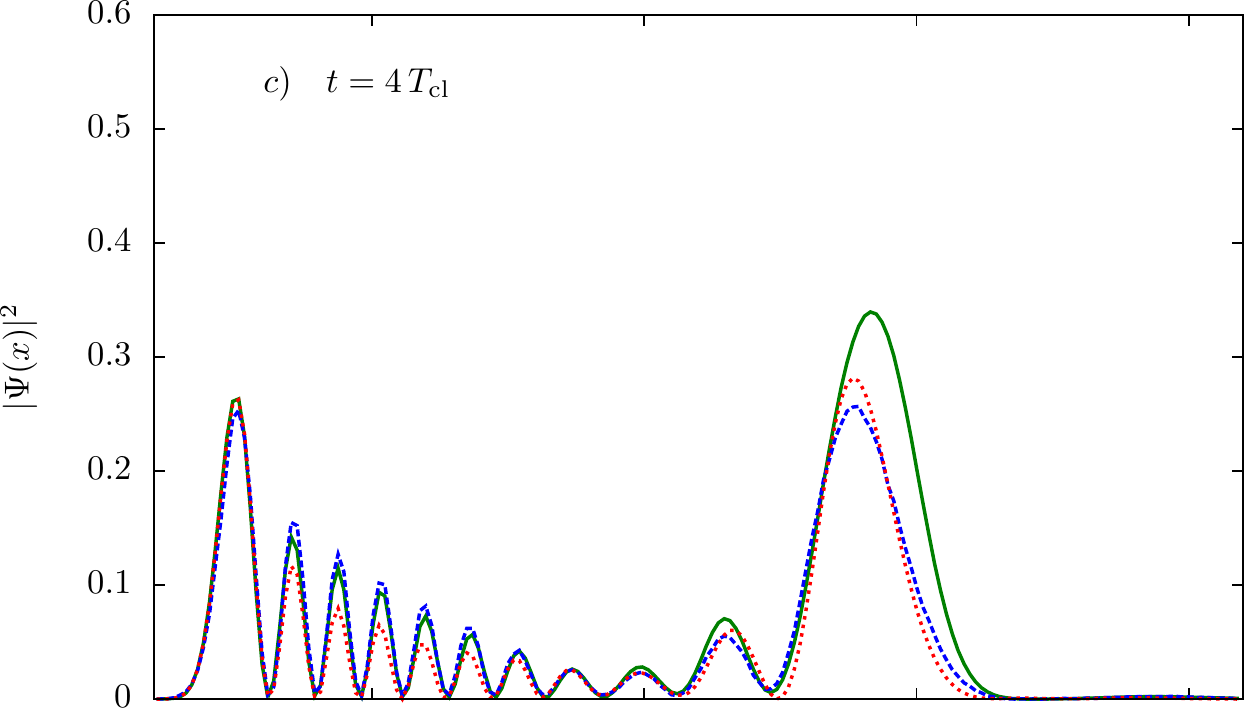}
  \end{minipage}
  \begin{minipage}[htp]{0.45\linewidth}
\includegraphics[width=\textwidth]{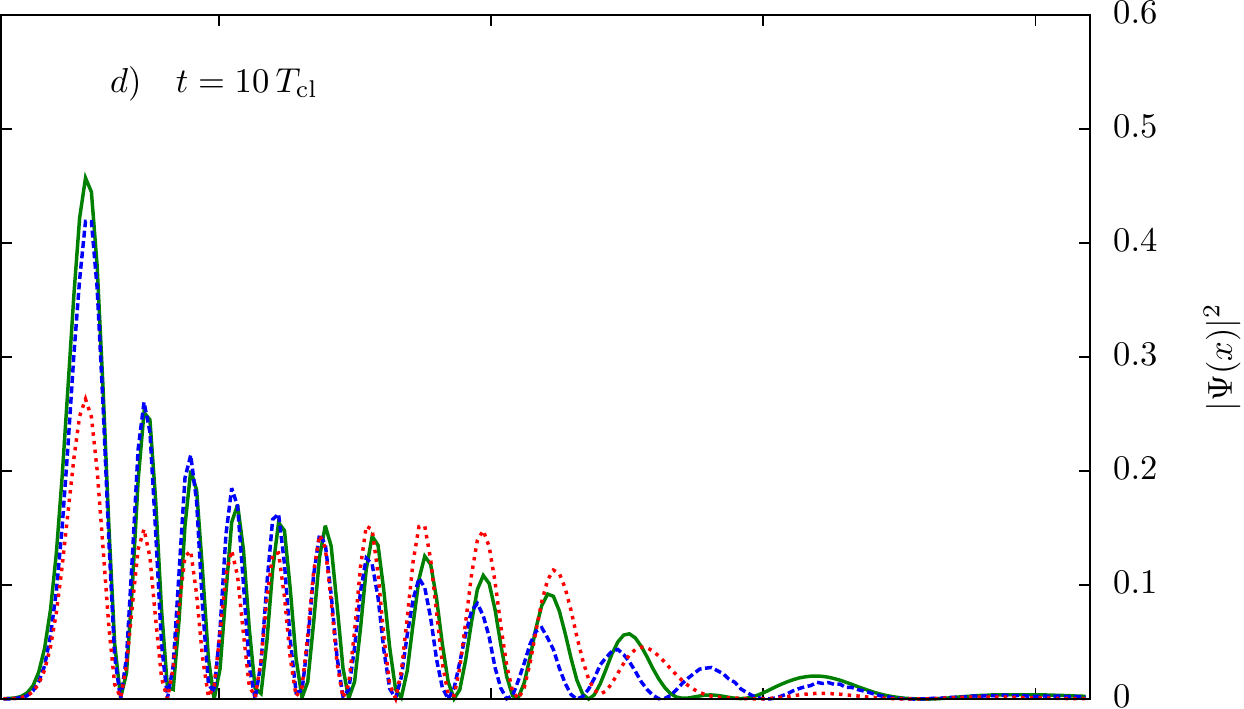}
  \end{minipage}

  \begin{minipage}[htp]{0.45\linewidth}
\includegraphics[width=\textwidth]{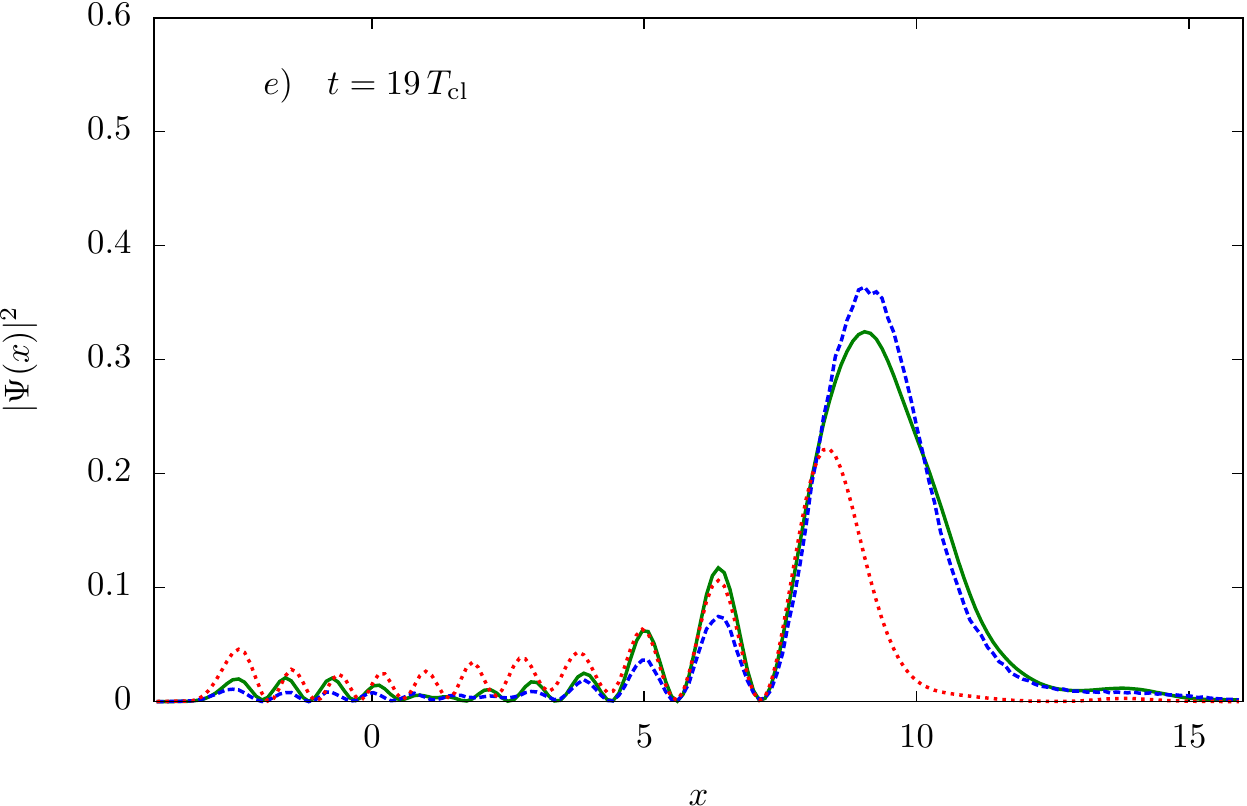}
  \end{minipage}
  \begin{minipage}[htp]{0.45\linewidth}
\includegraphics[width=\textwidth]{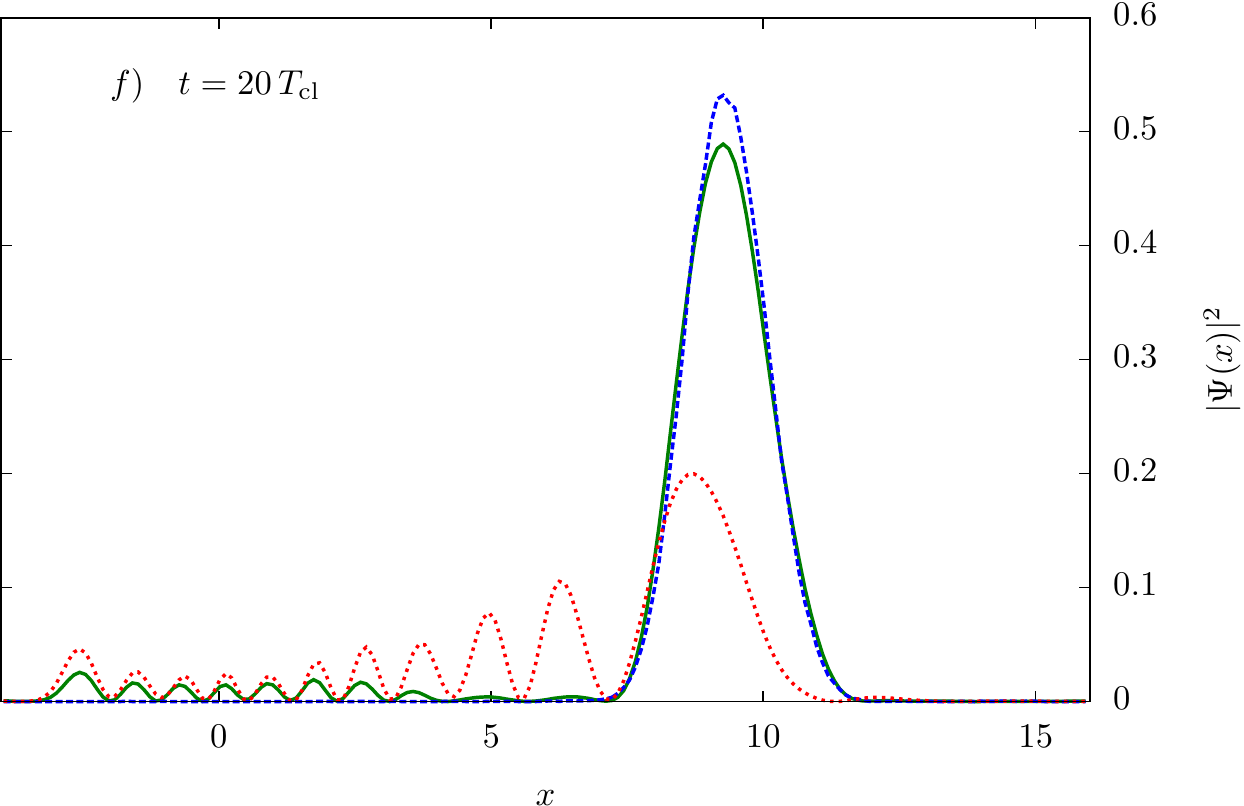}
  \end{minipage}
  \caption{\label{fig:psi_of_t}$|\Psi(x,t)|^2$ from FINCO reconstruction with complex time contours (green solid) and only with real time contours (red dotted) and split operator quantum propagation (blue dashed) for a range of times.
Panels a)-f) correspond to 0.5,1,4,10,19,20 classical periods.
Up to 120000 trajectories were used to reconstruct the wave functions for the longest time.
At $t=0.5T_{\rm{cl}}$, complex time propagation is not necessary at all.
To reach later times, complex time contours are required for accurate reconstruction and purely real contours can not access significant regions of coordinate space.
}
\end{figure}

Note that the results presented in Fig.~\ref{fig:psi_of_t} were obtained from complex time propagation in both cases.
For the ``purely real contour'' case, the trajectories were post-filtered to those that would in principle have been accessible via purely real time.
In practice such a purely real propagation is virtually impossible beyond $t=1T_{\rm{cl}}$ because a small but significant number of trajectories will pass too close to a dynamical singularity.
Even if these trajectories terminate on the correct side of this branching point, their action accumulates significant numerical noise.
Due to the exponential dependence of the final result on the action, these contributions taint the wave function reconstruction.
Filtering according to Sec.~\ref{sec:noise} is possible but would strongly depend on the filter limit $\epsilon$.
This sensitivity increases with propagation times, rendering a reasonable reconstruction as in Fig.~\ref{fig:psi_of_t}f) for $t=20T_{\rm{cl}}$ virtually impossible.

\bibliography{zotero_sj}

\begin{thebibliography}{10}
\expandafter\ifx\csname url\endcsname\relax
  \def\url#1{\texttt{#1}}\fi
\expandafter\ifx\csname urlprefix\endcsname\relax\def\urlprefix{URL }\fi
\expandafter\ifx\csname href\endcsname\relax
  \def\href#1#2{#2} \def\path#1{#1}\fi

\bibitem{herman_semiclasical_1984}
M.~F. Herman, E.~Kluk, A semiclasical justification for the use of
  non-spreading wavepackets in dynamics calculations, Chemical Physics 91~(1)
  (1984) 27--34.
\newblock \href {http://dx.doi.org/10.1016/0301-0104(84)80039-7}
  {\path{doi:10.1016/0301-0104(84)80039-7}}.

\bibitem{kay_semiclassical_2004}
K.~G. Kay, Semiclassical initial value treatments of atoms and molecules, Annu.
  Rev. Phys. Chem. 56~(1) (2004) 255--280.
\newblock \href {http://dx.doi.org/10.1146/annurev.physchem.56.092503.141257}
  {\path{doi:10.1146/annurev.physchem.56.092503.141257}}.

\bibitem{thoss_semiclassical_2004}
M.~Thoss, H.~Wang, Semiclassical {{Description}} of {{Molecular Dynamics
  Based}} on {{Initial}}-{{Value Representation Methods}}, Annual Review of
  Physical Chemistry 55~(1) (2004) 299--332.
\newblock \href {http://dx.doi.org/10.1146/annurev.physchem.55.091602.094429}
  {\path{doi:10.1146/annurev.physchem.55.091602.094429}}.

\bibitem{wu_nonadiabatic_2005}
Y.~Wu, M.~F. Herman, Nonadiabatic surface hopping {{Herman}}-{{Kluk}}
  semiclassical initial value representation method revisited: {{Applications}}
  to {{Tully}}'s three model systems, The Journal of Chemical Physics 123~(14)
  (2005) 144106.
\newblock \href {http://dx.doi.org/10.1063/1.2049251}
  {\path{doi:10.1063/1.2049251}}.

\bibitem{sun_semiclassical_1997}
X.~Sun, W.~H. Miller, Semiclassical initial value representation for
  electronically nonadiabatic molecular dynamics, The Journal of Chemical
  Physics 106~(15) (1997) 6346--6353.
\newblock \href {http://dx.doi.org/10.1063/1.473624}
  {\path{doi:10.1063/1.473624}}.

\bibitem{doll_complexvalued_1973}
J.~D. Doll, T.~F. George, W.~H. Miller, Complex-valued classical trajectories
  for reactive tunneling in three-dimensional collisions of {{H}} and {{H2}},
  The Journal of Chemical Physics 58~(4) (1973) 1343--1351.
\newblock \href {http://dx.doi.org/10.1063/1.1679366}
  {\path{doi:10.1063/1.1679366}}.

\bibitem{huber_generalized_1987}
D.~Huber, E.~J. Heller, Generalized {{Gaussian}} wave packet dynamics, The
  Journal of Chemical Physics 87~(9) (1987) 5302--5311.
\newblock \href {http://dx.doi.org/10.1063/1.453647}
  {\path{doi:10.1063/1.453647}}.

\bibitem{boiron_complex_1998}
M.~Boiron, M.~Lombardi, Complex trajectory method in semiclassical propagation
  of wave packets, The Journal of Chemical Physics 108~(9) (1998) 3431--3444.
\newblock \href {http://dx.doi.org/10.1063/1.475743}
  {\path{doi:10.1063/1.475743}}.

\bibitem{baranger_semiclassical_2001}
M.~Baranger, M.~A.~M. de~Aguiar, F.~Keck, H.~J. Korsch, B.~Schellhaa\ss{},
  Semiclassical approximations in phase space with coherent states, J. Phys. A:
  Math. Gen. 34~(36) (2001) 7227.
\newblock \href {http://dx.doi.org/10.1088/0305-4470/34/36/309}
  {\path{doi:10.1088/0305-4470/34/36/309}}.

\bibitem{goldfarb_bohmian_2006}
Y.~Goldfarb, I.~Degani, D.~J. Tannor, Bohmian mechanics with complex action:
  {{A}} new trajectory-based formulation of quantum mechanics, The Journal of
  Chemical Physics 125~(23) (2006) 231103.
\newblock \href {http://dx.doi.org/10.1063/1.2400851}
  {\path{doi:10.1063/1.2400851}}.

\bibitem{zamstein_non-adiabatic_2012}
N.~Zamstein, D.~J. Tannor, Non-adiabatic molecular dynamics with complex
  quantum trajectories. {{I}}. {{The}} diabatic representation, The Journal of
  Chemical Physics 137~(22) (2012) 22A517.
\newblock \href {http://dx.doi.org/10.1063/1.4739845}
  {\path{doi:10.1063/1.4739845}}.

\bibitem{zamstein_non-adiabatic_2012-1}
N.~Zamstein, D.~J. Tannor, Non-adiabatic molecular dynamics with complex
  quantum trajectories. {{II}}. {{The}} adiabatic representation, The Journal
  of Chemical Physics 137~(22) (2012) 22A518.
\newblock \href {http://dx.doi.org/10.1063/1.4739846}
  {\path{doi:10.1063/1.4739846}}.

\bibitem{zamstein_complex_2013}
N.~Zamstein, Complex quantum trajectories: {{Methodology}} development and
  chemical applications, Ph.D. thesis, Weizmann Institute of Science (2013).

\bibitem{zamstein_communication:_2014}
N.~Zamstein, D.~J. Tannor, Communication: {{Overcoming}} the root search
  problem in complex quantum trajectory calculations, The Journal of Chemical
  Physics 140~(4) (2014) 041105.
\newblock \href {http://dx.doi.org/10.1063/1.4862898}
  {\path{doi:10.1063/1.4862898}}.

\bibitem{huber_generalized_1988}
D.~Huber, E.~J. Heller, R.~G. Littlejohn, Generalized {{Gaussian}} wave packet
  dynamics, {{Schr{\"o}dinger}} equation, and stationary phase approximation,
  The Journal of Chemical Physics 89~(4) (1988) 2003--2014.
\newblock \href {http://dx.doi.org/10.1063/1.455714}
  {\path{doi:10.1063/1.455714}}.

\bibitem{de_aguiar_initial_2010}
M.~A.~M. {de Aguiar}, S.~A. Vitiello, A.~Grigolo, An initial value
  representation for the coherent state propagator with complex trajectories,
  Chemical Physics 370~(1\textendash{}3) (2010) 42--50.
\newblock \href {http://dx.doi.org/10.1016/j.chemphys.2010.01.020}
  {\path{doi:10.1016/j.chemphys.2010.01.020}}.

\bibitem{goldfarb_interference_2007}
Y.~Goldfarb, D.~J. Tannor, Interference in {{Bohmian}} mechanics with complex
  action, The Journal of Chemical Physics 127~(16) (2007) 161101.
\newblock \href {http://dx.doi.org/10.1063/1.2794029}
  {\path{doi:10.1063/1.2794029}}.

\bibitem{pal_generalized_2016}
H.~Pal, M.~Vyas, S.~Tomsovic, Generalized {{Gaussian}} wave packet dynamics:
  {{Integrable}} and chaotic systems, Phys. Rev. E 93~(1) (2016) 012213.
\newblock \href {http://dx.doi.org/10.1103/PhysRevE.93.012213}
  {\path{doi:10.1103/PhysRevE.93.012213}}.

\bibitem{kay_time-dependent_2013}
K.~G. Kay, Time-dependent semiclassical tunneling through barriers, Phys. Rev.
  A 88~(1) (2013) 012122.
\newblock \href {http://dx.doi.org/10.1103/PhysRevA.88.012122}
  {\path{doi:10.1103/PhysRevA.88.012122}}.

\bibitem{petersen_wave_2015}
J.~Petersen, K.~G. Kay, Wave packet propagation across barriers by
  semiclassical initial value methods, The Journal of Chemical Physics 143~(1)
  (2015) 014107.
\newblock \href {http://dx.doi.org/10.1063/1.4923221}
  {\path{doi:10.1063/1.4923221}}.

\bibitem{suarez_barnes_semiclassical_1993}
I.~M. Suarez~Barnes, M.~Nauenberg, M.~Nockleby, S.~Tomsovic, Semiclassical
  theory of quantum propagation: {{The Coulomb}} potential, Phys. Rev. Lett.
  71~(13) (1993) 1961--1964.
\newblock \href {http://dx.doi.org/10.1103/PhysRevLett.71.1961}
  {\path{doi:10.1103/PhysRevLett.71.1961}}.

\bibitem{mallalieu_semiclassical_1994}
M.~Mallalieu, C.~R. Stroud, Semiclassical dynamics of circular-orbit
  {{Rydberg}} wave packets, Phys. Rev. A 49~(4) (1994) 2329--2339.
\newblock \href {http://dx.doi.org/10.1103/PhysRevA.49.2329}
  {\path{doi:10.1103/PhysRevA.49.2329}}.

\bibitem{littlejohn_semiclassical_1986}
R.~G. Littlejohn, The semiclassical evolution of wave packets, Physics Reports
  138~(4) (1986) 193--291.
\newblock \href {http://dx.doi.org/10.1016/0370-1573(86)90103-1}
  {\path{doi:10.1016/0370-1573(86)90103-1}}.

\bibitem{shudo_stokes_1996}
A.~Shudo, K.~S. Ikeda, Stokes {{Phenomenon}} in {{Chaotic Systems}}: {{Pruning
  Trees}} of {{Complex Paths}} with {{Principle}} of {{Exponential Dominance}},
  Phys. Rev. Lett. 76~(22) (1996) 4151--4154.
\newblock \href {http://dx.doi.org/10.1103/PhysRevLett.76.4151}
  {\path{doi:10.1103/PhysRevLett.76.4151}}.

\bibitem{gelabert_log-derivative_2000}
R.~Gelabert, X.~Gim{\'e}nez, M.~Thoss, H.~Wang, W.~H. Miller, A
  {{Log}}-{{Derivative Formulation}} of the {{Prefactor}} for the
  {{Semiclassical Herman}}-{{Kluk Propagator}}\textdagger{}, J. Phys. Chem. A
  104~(45) (2000) 10321--10327.
\newblock \href {http://dx.doi.org/10.1021/jp0012451}
  {\path{doi:10.1021/jp0012451}}.

\bibitem{mitric_laser-field-induced_2009}
R.~Mitri{\'c}, J.~Petersen, V.~Bona{\v c}i{\'c}-Kouteck{\'y},
  Laser-field-induced surface-hopping method for the simulation and control of
  ultrafast photodynamics, Phys. Rev. A 79~(5) (2009) 053416.
\newblock \href {http://dx.doi.org/10.1103/PhysRevA.79.053416}
  {\path{doi:10.1103/PhysRevA.79.053416}}.

\bibitem{richter_sharc:_2011}
M.~Richter, P.~Marquetand, J.~Gonz{\'a}lez-V{\'a}zquez, I.~Sola,
  L.~Gonz{\'a}lez, {{SHARC}}: Ab {{Initio Molecular Dynamics}} with {{Surface
  Hopping}} in the {{Adiabatic Representation Including Arbitrary Couplings}},
  J. Chem. Theory Comput. 7~(5) (2011) 1253--1258.
\newblock \href {http://dx.doi.org/10.1021/ct1007394}
  {\path{doi:10.1021/ct1007394}}.

\bibitem{van_de_sand_irregular_1999}
G.~{van de Sand}, J.~M. Rost, Irregular {{Orbits Generate Higher Harmonics}},
  Phys. Rev. Lett. 83~(3) (1999) 524--527.
\newblock \href {http://dx.doi.org/10.1103/PhysRevLett.83.524}
  {\path{doi:10.1103/PhysRevLett.83.524}}.

\bibitem{zagoya_dominant-interaction_2012}
C.~Zagoya, C.-M. Goletz, F.~Grossmann, J.-M. Rost, Dominant-interaction
  {{Hamiltonians}} for high-order-harmonic generation in laser-assisted
  collisions, Phys. Rev. A 85~(4) (2012) 041401.
\newblock \href {http://dx.doi.org/10.1103/PhysRevA.85.041401}
  {\path{doi:10.1103/PhysRevA.85.041401}}.

\bibitem{kaushal_nonadiabatic_2013}
J.~Kaushal, O.~Smirnova, Nonadiabatic {{Coulomb}} effects in strong-field
  ionization in circularly polarized laser fields, Phys. Rev. A 88~(1) (2013)
  013421.
\newblock \href {http://dx.doi.org/10.1103/PhysRevA.88.013421}
  {\path{doi:10.1103/PhysRevA.88.013421}}.

\bibitem{pedatzur_attosecond_2015}
O.~Pedatzur, G.~Orenstein, V.~Serbinenko, H.~Soifer, B.~D. Bruner, A.~J. Uzan,
  D.~S. Brambila, A.~G. Harvey, L.~Torlina, F.~Morales, O.~Smirnova,
  N.~Dudovich, Attosecond tunnelling interferometry, Nat Phys 11~(10) (2015)
  815--819.
\newblock \href {http://dx.doi.org/10.1038/nphys3436}
  {\path{doi:10.1038/nphys3436}}.

\bibitem{kay_jacobian_2013}
K.~G. Kay, Jacobian for the {{FINCO}} formula, private communication (Dec.
  2013).

\bibitem{heller_classical_1976}
E.~J. Heller, Classical {{S}}-matrix limit of wave packet dynamics, The Journal
  of Chemical Physics 65~(11) (1976) 4979--4989.
\newblock \href {http://dx.doi.org/10.1063/1.432974}
  {\path{doi:10.1063/1.432974}}.

\end{thebibliography}
\bibliographystyle{elsarticle-num}

\end{document}